\documentclass[10pt,conference]{IEEEtran}
\IEEEoverridecommandlockouts
\usepackage{cite}
\usepackage{amsmath,amssymb,amsfonts}
\usepackage{multirow}
\usepackage{tcolorbox}
\usepackage{booktabs} 
\usepackage{graphicx}
\usepackage{textcomp}
\usepackage{xcolor}
\usepackage[linesnumbered,ruled,vlined]{algorithm2e}
\def\BibTeX{{\rm B\kern-.05em{\sc i\kern-.025em b}\kern-.08em
    T\kern-.1667em\lower.7ex\hbox{E}\kern-.125emX}}
\begin{document}

\title{TransferFuzz: Fuzzing with Historical Trace for Verifying Propagated Vulnerability Code}

\author{\IEEEauthorblockN{Siyuan Li\textsuperscript{1,2}, Yuekang Li\textsuperscript{3}, Zuxin Chen\textsuperscript{1,2}, Chaopeng Dong\textsuperscript{1,2}, Yongpan Wang\textsuperscript{1,2}\\
Hong Li\textsuperscript{1,2,*}, Yongle Chen\textsuperscript{4}, Hongsong Zhu\textsuperscript{1,2}}
\IEEEauthorblockA{\textit{\textsuperscript{1}Institute of Information Engineering, Chinese Academy of Sciences, Beijing, China} \\
\textit{\textsuperscript{2}School of Cyber Security, University of Chinese Academy of Sciences, Beijing, China}\\
\textit{\textsuperscript{3}School of Computer Science and Engineering, The University of New South Wales, Sydney, Australia}\\
\textit{\textsuperscript{4}College of Computer Science and Technology, Taiyuan University of Technology, Shanxi, China}\\
 \\
lisiyuan@iie.ac.cn, yuekang.li@unsw.edu.au, \{chenzuxin,dongchaopeng,wangyongpan,lihong\}@iie.ac.cn,\\
chenyongle@tyut.edu.cn,
zhuhongsong@iie.ac.cn}
 \thanks{* corresponding author: lihong@iie.ac.cn}}


\maketitle

\begin{abstract}
Code reuse in software development frequently facilitates the spread of vulnerabilities, making the scope of affected software in CVE reports imprecise. Traditional methods primarily focus on identifying reused vulnerability code within target software, yet they cannot verify if these vulnerabilities can be triggered in new software contexts. This limitation often results in false positives.
In this paper, we introduce TransferFuzz, a novel vulnerability verification framework, to verify whether vulnerabilities propagated through code reuse can be triggered in new software. 

Innovatively, we collected runtime information during the execution or fuzzing of the basic binary (the vulnerable binary detailed in CVE reports). This process allowed us to extract historical traces, which proved instrumental in guiding the fuzzing process for the target binary (the new binary that reused the vulnerable function). TransferFuzz introduces a unique Key Bytes Guided Mutation strategy and a Nested Simulated Annealing algorithm, which transfers these historical traces to implement trace-guided fuzzing on the target binary, facilitating the accurate and efficient verification of the propagated vulnerability.

Our evaluation, conducted on widely recognized datasets, shows that TransferFuzz can quickly validate vulnerabilities previously unverifiable with existing techniques. Its verification speed is 2.5 to 26.2 times faster than existing methods. Moreover, TransferFuzz has proven its effectiveness by expanding the impacted software scope for 15 vulnerabilities listed in CVE reports, increasing the number of affected binaries from 15 to 53. The datasets and source code used in this article are available at https://github.com/Siyuan-Li201/TransferFuzz.
\end{abstract}


\section{Introduction}

\newcounter{myCounter}

Code reuse is an important part of modern software development, providing benefits in terms of efficiency and functionality. A vast array of open-source code and third-party libraries are available on platforms like GitHub \cite{github}, SourceForge \cite{sourceforge} and Vcpkg \cite{vcpkg}. These resources enable developers to incorporate existing code into new projects, allowing them to concentrate on developing unique software features. Synopsys’ report \cite{Blackduck} in 2023 shows that 97\% of audit software reuses the code of at least one third-party library, highlighting the prevalence of code reuse in development practices.

Despite its benefits, code reuse can introduce significant risks. The widespread use of shared code can propagate vulnerabilities across numerous software projects \cite{LibAM}. While reused code benefits from iterative updates, it is not immune to flaws, which can spread as the code is reused. Security researchers typically report new vulnerabilities to platforms like CVE \cite{cve} or NVD \cite{nvd}. However, they may not fully consider the scope of the software affected by these vulnerabilities through code reuse. This paper mainly focuses on the vulnerability propagation verification of C/C++ binary, an area where the impact of vulnerabilities can extend beyond what is initially reported \cite{LibAM}.

Reflecting their distinct motivations, we categorize existing methods into three main types: Code Reuse Detection methods, Patch Presence Detection methods, and Directed Fuzzing methods. These methods can all detect the propagation of vulnerabilities due to code reuse.

Firstly, Code Reuse Detection methods aim to detect the presence of vulnerable functions in the target software. OSSPolice \cite{osspolice}, B2SFinder \cite{b2sfinder} and FirmSec \cite{firmsec} use constant features such as strings, arrays, and jump tables to detect code reuse and correlate the vulnerabilities. Besides, Gemini \cite{gemini}, JTrans \cite{jtrans}, and Asteria \cite{asteria} perform function similarity matching to detect vulnerabilities. Moreover, LibAM \cite{LibAM}, LibDI \cite{libdi}, and TPLite \cite{tplite} use code areas or function ratios from third-party libraries to detect propagated vulnerabilities. All these approaches focus on detecting the presence of the vulnerable code in the target software and don't further confirm whether the vulnerability poses a threat (\textbf{P1}). For instance, these vulnerable functions may have been patched.

Secondly, Patch Presence Detection methods aim to detect whether the reused vulnerability functions in the target software are patched. Fiber \cite{fiber} and PDiff \cite{pdiff} use symbolic execution methods to detect whether the code in the target software is closer to the vulnerable code or the patched code. LibvDiff \cite{libvdiff} uses function call paths to detect fine-grained versions of the target software, thus confirming whether the reused code is a patched version. However, recent research shows that both of these methods only detect whether the vulnerability is patched rather than judge whether the vulnerability can be triggered in the new software context \cite{robin}. This discrepancy has led to a large number of false positives\cite{LibAM} \textbf{(P1)}. There are two main reasons: the vulnerable function may be unreachable and the critical variables may be uncontrollable, which is described in detail in Section \setcounter{myCounter}{2}\Roman{myCounter}.

Thirdly, some researchers wanted to use directed fuzzing \cite{aflgo, uafuzz, beacon, WindRanger} for Patch Testing and Crash Production in software. However, the performance of current directed fuzzing approaches has been less than optimal. They often require several hours to verify a single vulnerability and struggle to trigger vulnerabilities necessitating complex logic \cite{uafuzz}\textbf{ (P2)}. To our knowledge, there is only one work OCTOPOCs \cite{octopocs} that uses historical knowledge from the Proofs-of-Concept (POCs) of third-party libraries like us and tries to trigger vulnerabilities in the new software. However, OCTOPOCs exhibit certain limitations. It is only suitable for small binaries to avoid path explosion and requires a specific input file for the target binary consistent with the basic binary \textbf{(P3)}. These limitations are discussed in detail in Section \setcounter{myCounter}{2}\Roman{myCounter}.


In summary, the existing methods suffer from the following problems:
\begin{itemize}
\item \textbf{P1}: Code Reuse Detection and Patch Presence Detection methods fail to confirm that the vulnerability is triggerable in the target software, thus leading to false positives.
\item \textbf{P2}: Directly using directed fuzzing techniques to verify vulnerabilities is slow and difficult to trigger vulnerabilities with complex logic.
\item \textbf{P3}: OCTOPOCs is limited by specific types of target binaries or input files, and cannot verify vulnerabilities in large binaries.
\end{itemize}

To overcome these challenges, we propose TransferFuzz, a novel vulnerability verification framework. We found that by collecting some runtime information of the \textbf{basic binary} (the vulnerable binary detailed in CVE reports), we can extract the historical traces to guide the fuzzing process for the \textbf{target binary} (new binary that reused the vulnerable function of basic binary). By using historical traces, we can achieve faster triggering of vulnerabilities in new software and trigger complex logic vulnerabilities (for \textbf{P2}). By generating POCs for new software, it is possible to automatically verify the vulnerability is an actual threat to the new software (for \textbf{P1} and \textbf{P3}).

TransferFuzz is not a new directed fuzzing technique, but a general framework designed to enhance existing directed fuzzing. Our framework utilizes runtime features to generate two types of historical traces: function-level traces and key-bytes traces. 
Different from Directed Fuzzing, we proposed a \textbf{Trace-guided Fuzzing} framework that incorporates a Key Bytes Guided Mutation strategy and a Nested Simulated Annealing algorithm to fuzz the target binary. This innovative approach allows for more accurate and efficient vulnerability verification. The motivation is detailed in Section \setcounter{myCounter}{2}\Roman{myCounter}.

Our evaluation, leveraging widely utilized datasets introduced by AFLGo and OCTOPOCs, demonstrates the capability of TransferFuzz to validate vulnerabilities that were previously unverifiable through existing techniques, achieving a verification speed that is 2.5 to 26.2 times faster than existing methods. Furthermore, TransferFuzz expands the impacted software scope of 15 vulnerabilities in the CVE report from 15 binaries to 53, underscoring its effectiveness in verifying propagated vulnerability code.

The contributions of this paper are as follows: 
\begin{itemize}
\item We propose TransferFuzz, the first vulnerability verification framework based on fuzzing to accurately and efficiently verify propagated vulnerability code.
\item TransferFuzz provides strong evidence for vulnerability propagation by utilizing the historical traces and triggering the vulnerability. 
\item We analyzed the software scope impacted by the 15 vulnerabilities in the CVE report and identified 38 additional affected binaries not previously reported.
\end{itemize}


\begin{figure}
\centering
    \includegraphics[width=\linewidth]{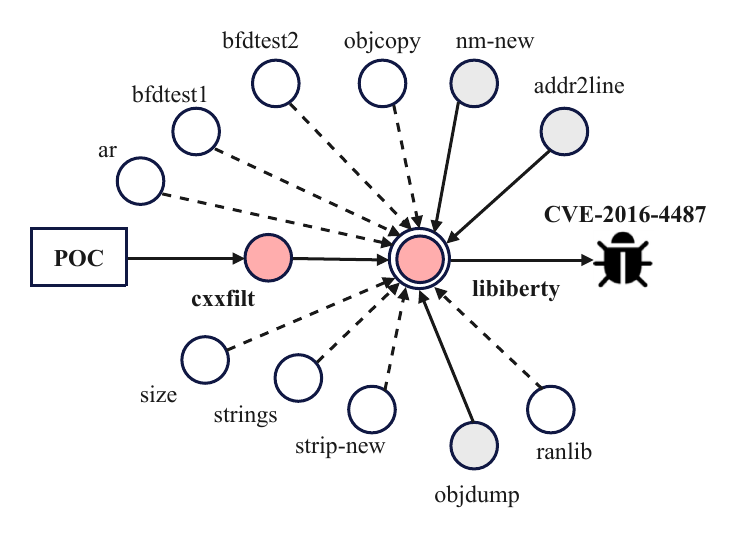}

\vspace{-1.0em}
\caption{Propagated vulnerability code. Red nodes are vulnerability information in CVE reports and historical research, white nodes are binaries that reuse vulnerable code, and gray nodes are binaries affected by the vulnerability.}
\vspace{-1.0em}
\end{figure}
\section{Motivation}


\begin{figure*}
\centering
    \includegraphics[width=\linewidth]{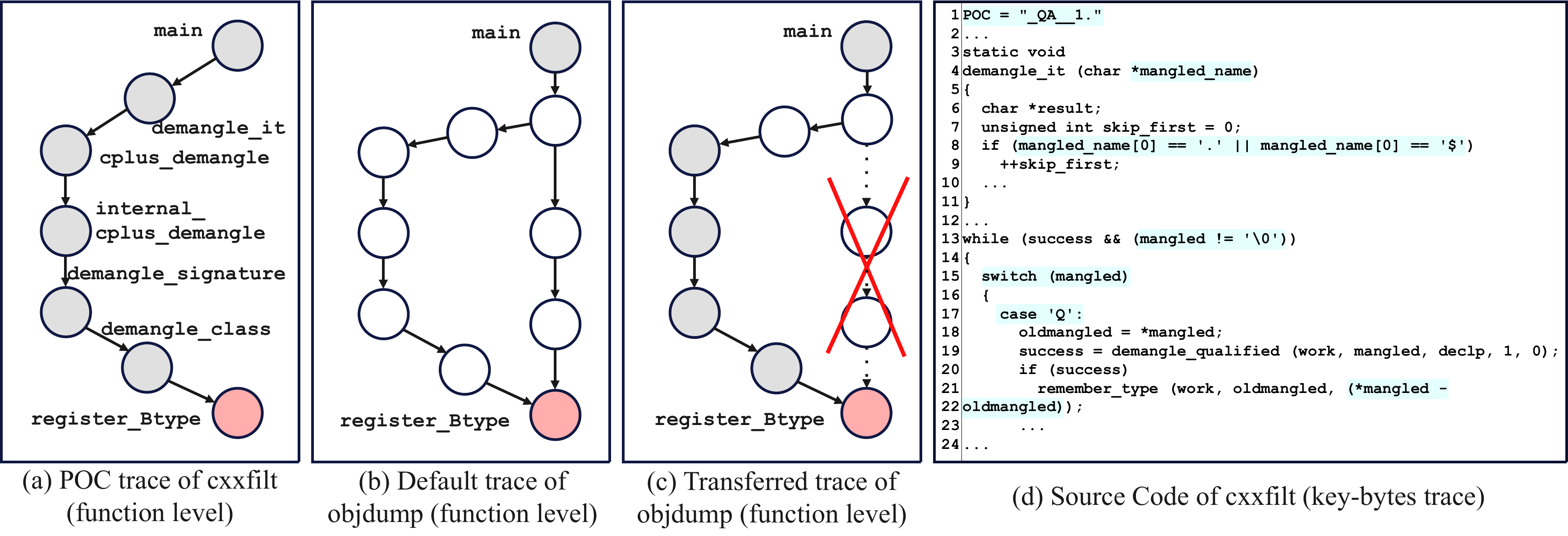}

\vspace{-1.0em}
\caption{A motivation example. The red node is the vulnerability function of CVE-2016-4487. Gray nodes are functions known to be on the vulnerability-triggering path, and white nodes are functions that may pass through. The blue code in 1(d) is the code in \textit{cxxfilt} that accesses POC related bytes}
\vspace{-1.0em}
\end{figure*}


We illustrate our approach with several examples. Figure 1 depicts the vulnerability CVE-2016-4487 in \textit{libiberty} as described in CVE \cite{cve} and NVD \cite{nvd} reports. AFLGo \cite{aflgo} employs fuzzing on \textit{cxxfilt} and generates a Proof of Concept (POC) that triggered this vulnerability. Subsequent analyses \cite{WindRanger, dafl} assumed that CVE-2016-4487 exclusively affected the \textit{cxxfilt} binary, ignoring potential impacts on other binaries. After manual analysis, we discovered that 11 new binaries reuse the vulnerable \textit{libiberty} code. However, since the input file formats required by these binaries are different, their POC cannot be universal, and it is difficult to determine which binaries can trigger this vulnerability. By manually constructing appropriate inputs, we successfully triggered the vulnerability in \textit{objdump}, \textit{nm-new}, and \textit{addr2line}. This example demonstrates that vulnerabilities in C/C++ binaries can propagate through code reuse, and the scope of software affected by a vulnerability extends beyond what vulnerability reports or prior research suggest.

However, not every instance of reused vulnerability code results in vulnerability propagation. As depicted in Figure 1, eight binaries remained unaffected by CVE-2016-4487. Despite containing the vulnerable code, the vulnerable functions were not reachable within those binaries. Expect the reachable, another factor also plays an important role. For instance, CVE-2016-10095 describes a stack-based buffer overflow vulnerability in \textit{LibTIFF}. When the value of the second parameter 'tag' of the function \textit{\_TIFFVGetField} is 0x13d, it will cause a buffer overflow. Although \textit{OpenJPEG} incorporates the vulnerable \textit{LibTIFF} code, the vulnerability does not threaten \textit{OpenJPEG} because the critical variable 'tag' is hard-coded, rendering the vulnerability's critical variables uncontrollable. Both Code Reuse Detection methods and Path Presentation Detection methods fail to confirm if the vulnerability functions are reachable or if critical variables are controllable (\textbf{P1}).

Besides, in our evaluation of \textit{objdump}, we observed that handling large binaries could lead to path explosion, causing methods like OCTOPOCs \cite{octopocs} to fail. Another challenge arises with vulnerabilities like CVE-2017-8393, an out-of-bounds read issue. OCTOPOCs struggles with scenarios where the crash cause does not align with specific bytes in a POC. In fact, OCTOPOCs can only handle special scenarios where the inputs to the basic software and the target software contain the full crash-cause byte and are processed in the same way (\textbf{P3}). It is worth noting that most of software in its dataset takes PDF files or ZIP files as input \cite{octopocs}.

\textbf{Our Approach} 
FuzzGuard \cite{fuzzguard} explores the use of historical fuzzing data within the same software to guide new fuzzing, and VulScope \cite{vulscope} explores the use of historical fuzzing data from different versions of the software to guide new fuzzing. Inspired by these two efforts, we propose leveraging historical fuzzing data from different software that share reused code areas to guide fuzzing (for \textbf{P2}). Our novel approach involves executing POCs from the basic binary to obtain historical traces that guide the fuzzing process in a new target binary to trigger the vulnerability (for \textbf{P1} and \textbf{P3}). Specifically, we focus on two types of historical traces: function-level traces and key-bytes traces. 

\textbf{Function-level Traces}: As demonstrated in Figure 2(a), executing the \textit{cxxfilt} for CVE-2016-4487 with POC reveals the function call sequence that triggers the vulnerability, which is treated as function-level traces. This function-level trace allows us to move beyond the blind fuzzing for \textit{objdump} depicted in Figure 2(b), adopting a more targeted approach as shown in Figure 2(c). This method only focuses on paths relevant to the vulnerability, thus speeding up the fuzzing process. Through static analysis, we found that there are 3150 different function calling paths in \textit{objdump} from the main function to the vulnerability function. However, we identified 42 paths by function-level traces of \textit{cxxfilt} and discarded the rest. 

\textbf{Key-bytes Traces}: Figure 2(d) shows how certain key bytes in the \textit{cxxfilt} POC can help bypass branch condition constraints in reused code (The highlighted code in the figure). These branches are also included in the target binary's reused code, hindering fuzzing speed. Extracting these bytes as key-bytes traces can significantly accelerate the fuzzing process, enhancing the efficiency of vulnerability verification.

\section{Methodology}
\subsection{Overview}
The workflow of TransferFuzz, presented in Figure 3, consists of three modules: Function-level Traces Extraction, Key-bytes Traces Extraction, and Trace Guided Fuzzing. We assume that Code Reuse Detection methods such as LibAM \cite{LibAM} have been used to detect that the target binary reuses the vulnerable code in the basic binary, and we want to further verify that the vulnerability can be triggered in the target binary. We take a basic binary with its POC and the target binary as inputs. Then, TransferFuzz judges whether the target binary is impacted by the vulnerability of the basic binary and generates a POC for the target binary if it is impacted.

\begin{figure*}
\centering
    \includegraphics[width=\linewidth]{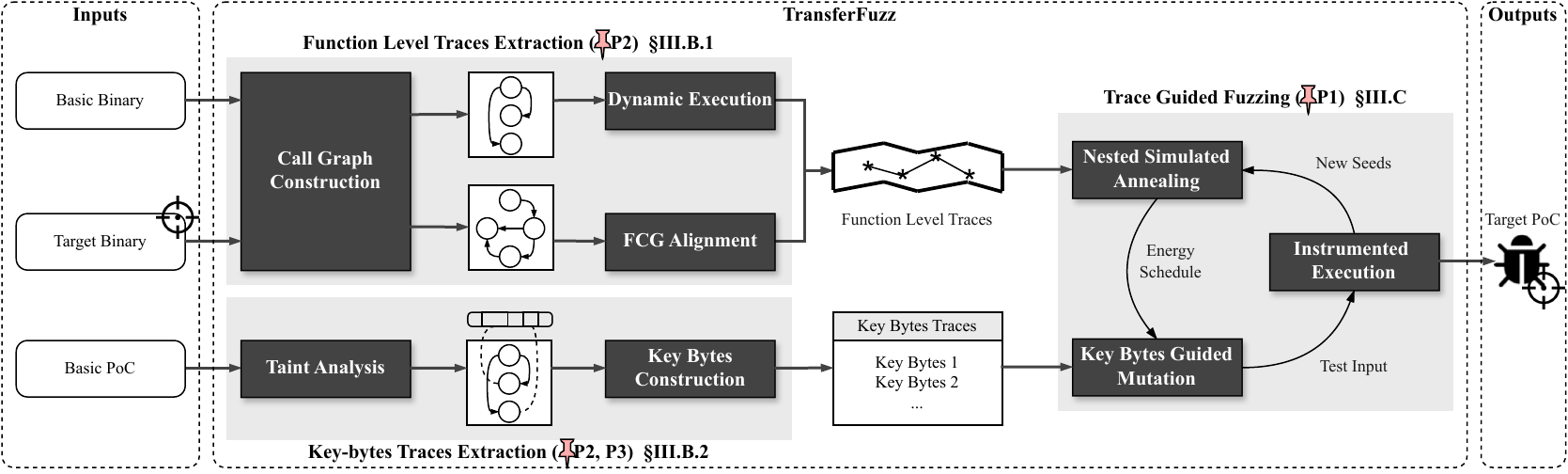}

\vspace{-1.0em}
\caption{The workflow of TransferFuzz.}
\vspace{-1.0em}
\end{figure*}

In the Function-level Traces Extraction module, we extract Function Call Sequence (FCS) from basic binary. Firstly, we execute the corresponding POC on the basic binary to collect run-time FCS as function-level traces to guide the subsequent fuzzing of the target binary. In addition, we find the path that triggers a vulnerability may not be single. Therefore, we also run a generic directed fuzz on the basic binary to collect more paths that may trigger the vulnerability. All the paths with run-time information that triggered the vulnerability are function-level traces and are taken as input for the final step. 

In the Key-bytes Traces Extraction module, we use taint analysis technology to extract the bytes in the POC that directly affect the conditional judgment statements in the reused code. These bytes are then conducted as a dictionary to guide the mutation of the target binary. In this way, some complex branch constraints can be bypassed when fuzzing the target binary.

In the Trace Guided Fuzzing module, we transfer historical traces to the target software to efficiently fuzz and thus verify vulnerabilities. During the Fuzzing process, in addition to the general Fuzzing mutation strategy, we propose the Key Bytes Guided Mutation strategy to use key-bytes traces to mutate seeds. Besides, we propose a Nested Simulated Annealing algorithm. In detail, we record the key jump points of vulnerability paths in function-level traces to form a state machine. For each test case, we schedule energy based on its execution state and the current most desired state to explore. Our goal is to continuously explore the most desired state path and avoid missing other possible paths as much as possible.

Finally, there are three possible results: the vulnerability is triggered, the reuse code is reached but the vulnerability is not triggered, and the reuse code area is not reached. For the first result, we consider the propagated vulnerability confirmed. For the latter two results, we consider that the vulnerability cannot be or is difficult to be triggered in the target software, and further manual verification is required. Compared with existing methods, although TransferFuzz has trouble making judgments about target binary without a crash, TransferFuzz can provide strong evidence for the propagation of the vulnerability that the POC triggers the vulnerability. The experimental results also further prove the vulnerability verification capability of TransferFuzz.

\subsection{Historical Trace Extraction}

In this section, our goal is to extract historical traces from basic binary. We found two types of runtime information that help a lot in fuzzing the target binary. One is the function call sequence (FCS) for the basic binary, the other is key bytes in POC. 

\subsubsection{\textbf{Function-level Traces Extraction}}

Firstly, we aim to extract function call sequence (FCS) from basic binary. Target binary and basic binary can often trigger vulnerabilities along the same path due to having the same reused code. Therefore, we would like to collect the vulnerability-triggering path of the basic binary and have the target binary fuzzing along this path. We have used two ways to extract FCS together. The first is to directly use the basic binary to execute POC to record run-time information, and the second is to use generic directed fuzzing on the basic binary to generate new POCs and record run-time information.

\textbf{Directly Execute POC}: For the basic binary whose vulnerability information with POCs, it can be directly executed POC to collect run-time information. Directly executing is very helpful when the vulnerability has the execution of complex logic and is not well triggered through generic fuzzing. We want to collect the function call sequence that triggered the vulnerability, and we want the target binary to run along these sequences during the fuzzing process, thus excluding irrelevant paths and enhancing the process's focus and efficiency. We run the basic binary and log the function call stack when the basic binary crashes. 

\textbf{Fuzzing Basic Binary}: POCs can be generated using generic directed fuzzing for basic binary without POCs in the vulnerability information. Besides, the basic binary with POCs can be expanded with more possible paths. In fact, using only one vulnerability triggering path may make TransferFuzz's function-level traces limited. In extreme cases, if the paths in the function-level traces are not the most easily triggered ones, it will affect the fuzzing of the target binary instead. Therefore, it is necessary to collect sufficiently rich paths for the basic binary by employing fuzzing in advance. This is also an offline process and does not affect the vulnerability verification efficiency of TransferFuzz because it only needs to be executed once on the basic binary to detect all possible propagation of the target binary. We use the state-of-the-art directed fuzzing method Windranger to perform directed fuzzing on the basic binary for crash reproduction, which is one of the tasks that directed fuzzing specializes in. By setting the source code line where the vulnerability is located, Fuzzer can calculate the distance from the current test case to the target code line, and guide the seed to mutate in the direction close to the vulnerability.

We do this based on such a motivation that the basic binary and target binary tend to have the same vulnerability triggering path in the reused code area due to having the same code. The vulnerability paths that are easy to reach during Fuzzing in the basic binary are also easier to reach in the target binary.

\textbf{FCG Alignment}: Finally, we need to align the basic binary's function-level traces with the target binary. We found that many target binaries make minor changes to the reused code. Therefore, we need to align the functions in the basic binary with those in the target binary using function matching. By doing this, the function-level traces can be directly applied to the target binary.

The easiest way to do this is to use the function names directly. This works when the target binary is not stripped. However, in practical scenarios, where some binary may be considered strip to remove function names, we can use Binary Code Similarity Detection (BCSD) methods for function matching. We used LibAM \cite{LibAM} in our evaluation to perform function matching when function names were missing after stripping. This step can be easily replaced by the existing BCSD methods \cite{jtrans, asteria}.

After the function matching, we can obtain a sub-path of the path in the history function call sequence of the basic binary. The original path reaches the vulnerability function from the main function of the basic binary, while the sub-path may be a path in the target binary that starts from one of the nodes in the original path and reaches the vulnerability function. We record the sub-path, which can directly guide the target binary's fuzzing.

\subsubsection{\textbf{Key-bytes Traces Extraction}}

In addition to function call sequences, we found that some bytes of the POC may affect key condition variables in the code. If these key bytes in the basic binary are extracted, it can help the fuzzing process of the target binary. This idea is similar to that of OCTOPOCs, but it is worth noting that OCTOPOCs just extract the key bytes in the POC and splice them together to generate a new POC, which is easy to fail. Instead, we hope to use these key bytes as a dictionary to assist the target binary's mutation process.

\begin{algorithm}[t]
\caption{Key Bytes Extraction Algorithm}

\KwIn{binaryExecutable $B$, reusedFunctionList $F$, proofOfConceptFile $P$}
\KwOut{accessedBytes $A$}

\SetKwFunction{FMain}{ExtractKeyBytes}
\SetKwProg{Fn}{Function}{:}{}
\Fn{\FMain{$B$, $F$, $P$}}{
    $PBytes \gets$ ReadBytes($P$)\;
    $BImages \gets$ LoadImages($B$)\;
    \ForEach{$image \in BImages$}{
        \ForEach{$func \in F$}{
            \If{$func$ is in $image$}{
                InstrumentEntryAndExit($func$)\;
            }
        }
    }
    $A \gets \emptyset$\;
    \For{each instruction $I$ in $B$}{
        \If{$I$ is MemoryAccess \textbf{and} IsInFunctionList($I$, $F$)}{
            $address \gets$ GetAccessAddress($I$)\;
            \If{$address$ corresponds to $PBytes$}{
                $byte \gets$ GetByteAt($address$, $PBytes$)\;
                $A \gets A \cup \{(address, byte)\}$\;
            }
        }
    }
    \KwRet $A$\;
}

\end{algorithm}

We implement this process using taint analysis techniques. Specifically, we mark the POC of the basic binary as tainted, and during execution, we pay attention to whether the comparison variables of the conditional statements in the reused code area come from the POC. Then, we compare the bytes accessed by reused code with bytes in POC to extract the bytes that are directly passed unchanged from the POC as key-bytes traces. The specific process is detailed in Algorithm 1. During the target binary's fuzzing process, when these tainted bytes are inserted at specific locations, it is easy to satisfy the corresponding conditional judgment in the reused code.

\subsection{Trace Guided Fuzzing}

In this section, our goal is to transfer the extracted historical trace to the target binary and efficiently verify the propagated vulnerability code in the target binary using Fuzzer with the transferred traces.

\subsubsection{\textbf{Nested Simulated Annealing}}

Unlike directed fuzz, which sets only one target line of code, TransferFuzz wants to utilize function call sequences from historical trace to schedule more energy to seeds exploring along vulnerability paths and less energy to seeds on unrelated paths.



Initially, a unique state machine is established for every execution path, meticulously recording every function encountered along each path throughout the fuzzing process. To manage the complexity of historical traces, distinct state machines are assigned to individual paths. Within these state machines, each state signifies a function node, charting a sequential path from the entry function of the reused code to the function harboring the vulnerability. Energies are judiciously scheduled to test cases based on a comparison between their achieved states and the latest state documented by the corresponding state machine, prioritizing those deemed most relevant.

Our objective is to preferentially allocate energies to the most recently identified states, while avoiding the pitfall of exclusively advancing along the deepest path, to the neglect of other potentially vulnerable paths. To achieve this balanced approach, we adopt the simulated annealing algorithm, as delineated in AFLGo. Simulated Annealing (SA) \cite{sa} is inspired by the annealing process in metallurgy, a technique that involves heating and controlled cooling of a material to increase its crystal size and reduce its defects. In AFLGo, a designated moment, denoted as $tx$, signifies the transition from a phase of extensive exploration to a focused exploitation phase. Intuitively, at the juncture of $tx$, the behavior of the simulated annealing algorithm mirrors that of a classical gradient descent algorithm, effectively embodying a greedy search approach. Before reaching $tx$, the algorithm ensures a thorough exploration of alternative paths, thus maintaining a holistic assessment of potential vulnerabilities.


Unlike AFLGo's singular Markov chain approach, we model each execution path as an independent state machine, with each state node representing a function node on the path. For each state machine, we maintain a time list, documenting the initial encounter with each state. This approach converts the task of energy scheduling for each state into determining their joint probability distribution. Upon the emergence of a new state, we initially set the energy distribution between the new and existing states to 0.5. This ratio is dynamically adjusted via the simulated annealing algorithm, incrementally favoring the newer states with increased energy while reducing that of the older states. Such a strategy facilitates the individualized management of simulated annealing timelines for each state, thereby ensuring the state machine collectively gravitates towards the most recent state. The pseudo-code for the algorithm is given in Algorithm 2.






\begin{algorithm}[t]
\caption{Nested Simulated Annealing Algorithm}

\KwIn{current state $cur$, states list $states$, start time list $time$, current time $t$}
\KwOut{energy $E$}

\SetKwFunction{FCalcEnergy}{CalcEnergy}
\SetKwFunction{FSimulatedAnnealing}{SimulatedAnnealing}
\SetKwProg{Fn}{Function}{:}{}
\Fn{\FCalcEnergy{$cur$, $states$, $time$, $t$}}{
    \If{$cur$ is new}{
        NewState($t$, $states$, $time$)\;
        \KwRet $0.5$\;
    }
    $idx \gets$ Index of $cur$ in $states$\;
    $start \gets time[idx]$\;
    $temp \gets$ \FSimulatedAnnealing{$start$, $t$}\;
    $E \gets 1 - 0.5 \times temp$\;
    
    \For{$i = idx+1$ \KwTo length of $states$}{
        $newStart \gets time[i]$\;
        $newTemp \gets$ \FSimulatedAnnealing{$newStart$, $t$}\;
        $E \gets E \times 0.5 \times newTemp$\;
    }
    \KwRet $E$\;
}

\;

\SetKwProg{Fn}{Function}{:}{}
\Fn{\FSimulatedAnnealing{$start$, $t$}}{
    $T \gets$ Cooling($t, start$)\; 
    \KwRet $T$\;
}

\end{algorithm}






In detail, for each input seed, the energy is calculated based on the state reached during the execution. The calculation of energy for every state is defined as Equation 1:

\begin{equation}
E_j = 
\begin{cases} 
\prod\limits_{i>j}0.5T_i & \text{if } j=0, \\
(1-0.5T_j)\cdot\prod\limits_{i>j}0.5T_i &  \text{if } j>0.
\end{cases}
\end{equation}

Within this context, $E_j$ denotes the energy of the $j\ {th}$ state, while $T_j$ represents the temperature parameter \cite{aflgo} for the $j\ {th}$ state, calculated using the simulated annealing algorithm.

We present the detailed procedure of the nested simulated annealing algorithm using the example in Figure 4. Initially, TransferFuzz operates like a generic directed fuzzing tool before reaching the reused code area. At this stage, energy scheduling is a basic simulated annealing strategy \cite{aflgo}. However, once \textit{state0} of the state machine is encountered, we trigger the first simulated annealing algorithm. This algorithm allocates half of the energy (0.5) to seeds that have achieved \textit{state0} and distributes the remaining half to those that have not.  Over time, the energy for \textit{state0} escalates to 1, diminishing the energy for seeds failing to reach \textit{state0} to zero.  Upon the initial reach of the new \textit{state1}, the second simulated annealing algorithm kicks in, assigning half of the energy (0.5) to seeds at \textit{state1} while the other 0.5, including those stuck at \textit{state0}, share the rest. This 0.5 split still adheres to the principles of the first algorithm. A new simulated annealing algorithm is activated whenever a new state is reached. By using this nested simulated annealing algorithm, the energy of the seeds that are not related to the vulnerability paths in the historical trace decay rapidly and mutates in the direction that drives the state machine forward.

\subsubsection{\textbf{Key Bytes Guided Mutation}}
TransferFuzz adopts the mutation strategy of traditional directed fuzzing but enhances it significantly. TransferFuzz distinguishes itself by utilizing key bytes extracted from the binary's historical traces to guide test case generation. Specifically, each unique sequence of consecutively accessed bytes in memory is treated as a row in the dictionary. By systematically integrating these key bytes into a dictionary and applying it at every mutation phase, TransferFuzz can bypass some difficult-to-implement branch constraints and demonstrably boost fuzzing efficiency, as supported by our experimental results. 

Note that the KBGM algorithm, while similar to the taint analysis used in AFL++ \cite{aflplus}, Windranger \cite{WindRanger}, and REDQUEEN \cite{redqueen}, is more robust. It extracts knowledge from another binary and can handle branches with complex conditions. This capability allows it to handle complex vulnerabilities, as demonstrated in the evaluation.

\begin{figure}
\centering
    \includegraphics[width=\linewidth]{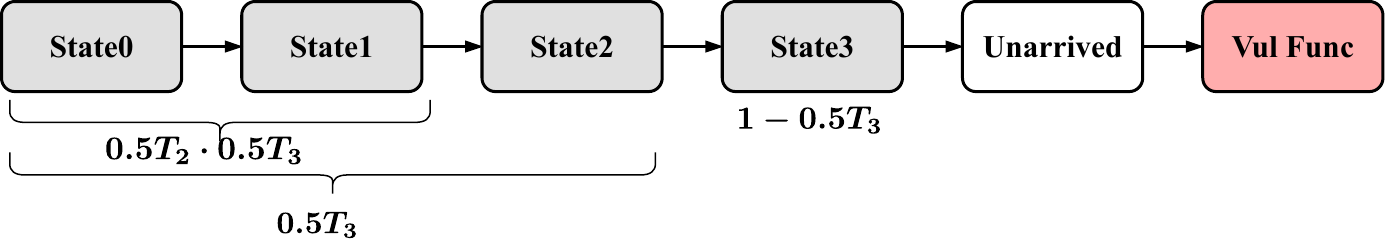}

\vspace{-1.0em}
\caption{State machine diagram in NSA algorithm.}
\vspace{-2.0em}
\end{figure}
\section{Implementation}

\begin{table*}
  \centering
  \caption{dataset scale}
  \label{tab:freq1}
  \begin{tabular}{cccccccc}
    \toprule
    Vul & Source & Type & CrashFunc & Code Reuse & Propogated & Propogated Binaries\\
    \midrule
    CVE-2016-10095 & tiffsplit & CWE-119 & \_TIFFVGetField & 24 & 2 &  thumbnail, tiffcmp\\
    CVE-2016-5318 & thumbnail & CWE-119 & \_TIFFVGetField & 24 & 2 &  tiffsplit, tiffcmp\\
    CVE-2016-4487 & cxxfilt & CWE-416 & register\_Btype & 11 & 3 &  objdump, nm-new, addr2line\\
    CVE-2016-4489 & cxxfilt & CWE-190 & string\_appendn & 11 & 3 &  objdump, nm-new, addr2line\\
    CVE-2016-4490 & cxxfilt & CWE-190 & d\_unqualified\_name & 11 & 3 &  objdump, nm-new, addr2line\\
    CVE-2016-4491 & cxxfilt & CWE-119 & d\_print\_comp\_inner & 11 & 3 &  objdump, nm-new, addr2line\\
    CVE-2016-4492 & cxxfilt & CWE-119 & do\_type & 11 & 3 &  objdump, nm-new, addr2line\\
    CVE-2016-6131 & cxxfilt & CWE-20 & demangle\_class\_name & 11 & 3 &  objdump, nm-new, addr2line\\
    CVE-2017-7303 & strip & CWE-125 &  section\_match & 11 & 1 & objcopy\\
    CVE-2017-11733 & swftocxx & CWE-125 &  stackswap & 6 & 4 & swftophp, swftoperl, swftopython, swftotcl\\
    CVE-2018-8807 & swftophp & CWE-416 &  getString & 6 & 4 & swftocxx, swftoperl, swftopython, swftotcl\\
    CVE-2018-8962 & swftophp & CWE-416 &  getName & 6 & 4 & swftocxx, swftoperl, swftopython, swftotcl\\
    CVE-2017-18267 & Poppler & CWE-476 &  FoFiType1C::getOp & 1 & 1 & xpdf\\
    CVE-2018-11102 & libav & CWE-119 &  mov\_probe & 1 & 1 & ffmpeg\\
    CVE-2018-20330 & libjpeg-turbo & CWE-190 &  \_\_memcpy\_avx\_unaligned & 1 & 1 & mozjpeg\\
  \bottomrule

\end{tabular}
  \vspace{-0.9em}
\end{table*}

We implement the TransferFuzz prototype with over 2000 lines of Python and C code. TransferFuzz is a generalized transferred trace-based fuzzing framework that can based on any of the directed fuzzing techniques. We have selected the latest SelectFuzz \cite{selectfuzz}, to implement TransferFuzz.

\textbf{Historical Trace Extraction} To extract function call sequences from basic binary, we use GDB \cite{gdb} for dynamic debugging, and we wrote Python scripts to automate calls to GDB to extract function call stacks after crashes. For the key bytes, we use PIN \cite{pin} for taint analysis and judge whether tainted bytes come directly from POC bytes.


\textbf{Trace Guided Fuzzing}
For the Key Bytes Guided Mutation strategy, we simply maintain the extracted key bytes as a dictionary and enter them as arguments when calling fuzzer. For the Nested Simulated Annealing algorithm, we save the function call sequences in the historical trace as several subscripts of a global array that records the test case coverage edges during Fuzzing. After executing a new coverage edge, we determine the state of the current test case by detecting whether the edge corresponds to the several subscripts of the global array or not. Additionally, we update the global state machine with this information. By doing so, we have implemented TrasnferFuzz with only negligible space and time complexity. This makes TransferFuzz no slower to execute than generic directed Fuzzing. Finally, we developed a shell script to automatically determine the completion of the verification process by checking the crash address via GDB.


\section{Evaluation}

In this section, we evaluate TransferFuzz and the existing state-of-the-art (SOTA) works. We provide a comprehensive analysis, including an investigation into the bad cases of existing works, and identify the reasons for their shortcomings. Before that, we describe our experimental setup, the dataset used, and the compared works.

First, in the experiments, we aim to answer the following research questions:

\textbf{RQ1}: How TransferFuzz performs in the Vulnerability Verification of Reused Code task?

\textbf{RQ2}: How efficiently TransferFuzz triggers the vulnerability in target binaries? 

\textbf{RQ3}: How does every component in TransferFuzz affect the overall performance?

Then, we detail the experimental setup. The system operates on Ubuntu 22.04, powered by an Intel Xeon CPU with 128 cores at 3.0GHz and hyperthreading capabilities. Each fuzzer is executed within a Docker container. 

Finally, we employed the following metrics to evaluate the performance of different methods:

To answer RQ1, we use Precision and Recall to evaluate whether TransferFuzz can verify more vulnerabilities and reduce false positives compared to existing methods. Specifically, $TP$ represents the number of correctly verified propagated vulnerabilities, while $TN$ denotes the number of non-propagated vulnerabilities that are correctly detected as non-propagated. Conversely, $FN$ indicates the number of non-propagated vulnerabilities that are incorrectly verified while $FP$ presents the number of propagated vulnerabilities that are incorrectly detected as propagated. Moreover, the equations of Precision and Recall are as follows:

 \begin{equation}
 Precision = \frac{TP}{TP + FP}
 \end{equation}

 \begin{equation}
 Recall = \frac{TP}{TP + FN}
 \end{equation}


To answer the RQ2 and RQ3, we use the Time-to-Exposure (TTE) to evaluate the efficiency. Similar to previous work \cite{WindRanger}, each method was repeated 10 times to take the average $\mu$TTE with a time budget of 12 hours. We calculate the sum of $\mu$TTE for vulnerabilities successfully triggered 10 times and evaluate the percent TrasferFuzz speed up.

\textbf{TTE}: Time-to-Exposure (TTE) is used to measure the time used by the fuzzer to trigger the vulnerability.



\subsection{Dataset}

We performed code reuse detection on the datasets from AFLGo \cite{aflgo}, SelectFuzz \cite{selectfuzz}, and DAFL \cite{dafl}, extracting binaries with propagated vulnerable functions. Additionally, although OCTOPOCs does not publish its dataset, we found six of its PDF processing software on the internet. In total, we collected 15 CVE vulnerabilities that propagate due to code reuse and 76 related binaries. The dataset is summarized in Table 1.

Using LibAM, the state-of-the-art code reuse detection technology, we identified 146 new potential vulnerabilities by checking for the presence of vulnerable code. After manually filtering out 108 false positives, we found that the initial 15 CVE vulnerabilities had broader implications, affecting more software and extending to 53 distinct vulnerabilities. This dataset demonstrates the capability of TransferFuzz in vulnerability validation and proof-of-concept (POC) generation.


\subsection{Compared Methods}
In this evaluation, we introduced the comparison between TransferFuzz and various existing methods. The comparison methods are as follows:

\textbf{LibAM}\cite{LibAM}: We employed LibAM, the SOTA third-party library code reuse detection method, as a representative of code reuse-based approaches.

\textbf{OCTOPOCs}\cite{octopocs}: OCTOPOCs is the only one that has a similar idea to ours, employing symbolic execution and taint analysis for the verification of propagating vulnerable code.

\textbf{AFLGo}\cite{aflgo}: AFLGo is the earliest and classic Directed Fuzzing technique.

\textbf{Windranger}\cite{WindRanger}: Windranger is a directed fuzzing technology that uses deviation basic blocks.

\textbf{SelectFuzz}\cite{selectfuzz}: SelectFuzz is an advanced directed fuzzing approach that employs selective instrumentation to enhance efficiency.

\textbf{DAFL}\cite{dafl}: DAFL is an advanced directed fuzzing technology that uses data flow distance metrics

\begin{table}
  \centering
  \caption{Accuracy in the Vulnerability
Verification task}
  \label{tab:freq3}
  \resizebox{\columnwidth}{!}{%
    \begin{tabular}{ccccccc}
      \toprule
      \textbf{Model} & \textbf{TP} & \textbf{FP} & \textbf{TN} & \textbf{FN} & \textbf{Precision} & \textbf{Recall}\\
      \midrule
      LibAM &  38 &  108 &  0 &  0 & 0.260 & 1.0\\
      {OCTOPOCs} &  15 &  0 &  108 &  23 & 1.0 & 0.395\\
      {AFLGo} & 27 &  0 &  108 &  11 & 1.0 & 0.711\\
      {Windranger} & 29 &  0 &  108 &  10 & 1.0 & 0.763\\
      {SelectFuzz} & 29 &  0 &  108 &  10 & 1.0 & 0.763\\
      {DAFL} & 20 &  0 &  108 &  18 & 1.0 & 0.526\\
      {TransferFuzz} & \textbf{38} &  \textbf{0} &  \textbf{108} &  \textbf{0} & \textbf{1.0} & \textbf{1.0}\\
      \bottomrule
    \end{tabular}%
  }
\end{table}

\begin{table*}
\centering
\caption{The number of runs and efficiency for each vulnerability.}
\label{table1}
\renewcommand{\arraystretch}{1.2} 
\begin{tabular}{cc|cc|cc|cc|cc|cc}
\toprule[2pt]
\multicolumn{1}{c}{\multirow{2}{*}{\textbf{No.}}} & \multicolumn{1}{c}{\multirow{2}{*}{\textbf{Vulnerability}}} & \multicolumn{2}{|c|}{\textbf{AFLGo}} & \multicolumn{2}{c|}{\textbf{Windranger}}  & \multicolumn{2}{c|}{\textbf{SelectFuzz}} & \multicolumn{2}{c|}{\textbf{DAFL}} & \multicolumn{2}{c}{\textbf{TransferFuzz}} \\
\cline{3-12}
{} & {} & Runs & $\mu$\textbf{TTE} & Runs & $\mu$\textbf{TTE} & Runs & $\mu$\textbf{TTE} & Runs & $\mu$\textbf{TTE} & Runs& $\mu$\textbf{TTE} \\
\hline
1 & thumbnail-2016-10095 & 10 & 10m35s & 10 & 3m27s  & 10 & 2m13s & 10 & 3m38s & 10 & \textbf{0m56s}\\
2 & tiffcmp-2016-10095 & 10 & 8m24s & 10 & 5m16s & 10 & 2m32s & 10 & 4m17s & 10 & \textbf{1m49s}\\
\hline
3 & tiffsplit-2016-5318 & 10 & 7m27 & 10 & 6m5s & 10 & 1m24s & 10 & 2m53s  & 10 & \textbf{1m28s}\\
4 & tiffcmp-2016-5318 & 10 & 7m12s & 10 & 4m39s & 10 & 3m5s & 10 & 4m24s & 10 & \textbf{1m38s}\\
\hline
5 & objdump-2016-4487 & 10  & 27m23s & 10 & 5m32s & 10 & 14m17s & 10 & 9m54s & 10  & \textbf{0m5s}\\
6 & nm-new-2016-4487 & 10  & 35m31s & 10 & 6m12s & 10 & 12m32s & 10 & 13m56s & 10 & \textbf{0m40s}\\
7 & addr2line-2016-4487 & 10  & 1h5m45s & 10 & 1h14m23s & 10 & 43m23s & 10  & 1h3m24s & 10 &  \textbf{0m38s}\\
\hline
8 & objdump-2016-4489 & 10  & 48m5s & 10 & 7m16s & 10 & 17m23s & 10 & 13m45s & 10 & \textbf{0m4s}\\
9 & nm-new-2016-4489 & 10  & 34m51s & 10 & 3m42s & 10 & 13m51s & 10 & 9m13s & 10 & \textbf{0m27s}\\
10 & addr2line-2016-4489 & 10  & 1h47m15s & 10 & 1h52m0s & 10 & 1h3m25s & 10 & 1h24m26s & 10 & \textbf{0m51s}\\
\hline
11 & objdump-2016-4490 & 10  & 21m45s & 10 & 4m14s & 10 & 13m26s & 10 & 9m42s & 10 & \textbf{0m4s}\\
12 & nm-new-2016-4490 & 10  & 35m23s & 10 & 11m24s & 10 & 9m52s & 10 & 13m26s & 10 & \textbf{1m3s}\\
13 & addr2line-2016-4490 & 10  & 2h4m21s & 10 & 1h3m25s & 10 & 58m25s & 10 & 42m52s & 10 & \textbf{0m53s}\\
\hline
14 & objdump-2016-4491 & 0 & N.A. & 0 & N.A. & 0 & N.A. & 0 & N.A. & 10 & \textbf{3m37s}\\
15 & nm-new-2016-4491 & 0 & N.A. & 0 & N.A. & 0 & N.A. & 0 & N.A. & 10 & \textbf{4m15s}\\
16 & addr2line-2016-4491 & 0 & N.A. & 0 & N.A. & 0 & N.A. & 0 & N.A. & 10 & \textbf{7m14s}\\
\hline
17 & objdump-2016-4492 & 10  & 56m24s & 10 & 46m11s & 10 & 23m15s & 10 & 35m1s & 10 & \textbf{0m32s}\\
18 & nm-new-2016-4492 & 10 & 1h2m42s & 10 & 58m15s & 10 & 44m35s & 10 & 39m25s & 10 & \textbf{0m40s}\\
19 & addr2line-2016-4492 & 0 & N.A. & 0 & N.A. & 3 & 2h46m & 2  & 3h41m58s & 10 &  \textbf{0m17s}\\
\hline
20 & objdump-2016-6131 & 0 & N.A. & 0 & N.A. & 0 & N.A. & 0 & N.A. & 10 & \textbf{0m4s}\\
21 & nm-new-2016-6131 & 0 & N.A. & 0 & N.A. & 0 & N.A. & 0 & N.A. & 10 & \textbf{0m56s}\\
22 & addr2line-2016-6131 & 0 & N.A. & 0 & N.A. & 0 & N.A. & 0 & N.A. & 10 & \textbf{0m32s}\\
\hline
23 & objcopy-2017-7303 & 0 & N.A. & 3 & 2h33m & 5 & 1h57m12s & 0 & N.A. & 6 & \textbf{1h30m2s}\\
\hline
24 & swftophp-2017-11733 & 10 & 5m16s & 10 & 1m59s & 10 & 0m32s & 10 & 1m23s & 10 & \textbf{0m6s}\\
25 & swftoperl-2017-11733 & 10 & 4m25s & 10 & 1m42s & 10 & \textbf{0m34s} & 10 & 2m0s & 10 & 0m52s\\
26 & swftopython-2017-11733 & 10 & 6m43s & 10 & 3m1s & 10 & 0m46s & 10 & 1m25s & 10 & \textbf{0m28s}\\
27 & swftotcl-2017-11733 & 10 & 3m25s & 10 & 2m48s & 10 & 1m22s & 10 & 2m34s & 10 & \textbf{0m25s}\\
\hline
28 & swftocxx-2018-8807 & 10 & 2h24m17s & 10 & 40m45s & 10  & 32m14s & 0 & N.A. & 10 & \textbf{20m2s}\\
29 & swftoperl-2018-8807 & 10 & 2h4m17s & 10 & 51m32s  & 10 & 19m15s & 0 & N.A. & 10 & \textbf{9m21s}\\
30 & swftopython-2018-8807 & 10 & 2h37m42s & 10 & 47m13s  & 10 & 26m41s & 0 & N.A. & 10 & \textbf{17m2s}\\
31 & swftotcl-2018-8807 & 10 & 2h8m32s & 10 & 54m21s & 10  & 29m1s & 0 & N.A. & 10 & \textbf{22m49s}\\
\hline
32 & swftocxx-2018-8962 & 10 & 3h43m54s & 10 &  1h13m45s & 10  & 45m15s& 0 & N.A. & 10 & \textbf{33m42s}\\
33 & swftoperl-2018-8962 & 10 & 3h24m16s & 10 & 56m34s & 10  & 33m25s & 0 & N.A. & 10 &  \textbf{29m15s} \\
34 & swftopython-2018-8962 & 10 & 3h52m1s & 10 & 1h3m52s & 10  & 58m21s  & 0 & N.A. & 10 &  \textbf{55m20s}\\
35 & swftotcl-2018-8962 & 10 & 3h35m15s & 10 & 52m16s & 10  & 1h1m15s & 0 & N.A. & 10 & \textbf{46m16s}\\

\bottomrule[1pt]
\multicolumn{2}{c|}{\multirow{2}{*}{\textbf{$\mu$TTE inc (all binaries)}}} & \multicolumn{2}{c|}{\multirow{2}{*}{\textbf{+850\%}}} & \multicolumn{2}{c|}{\multirow{2}{*}{\textbf{+344\%}}} & \multicolumn{2}{c|}{\multirow{2}{*}{\textbf{+248\%}}}& \multicolumn{2}{c|}{\multirow{2}{*}{\textbf{+2620\%}}}\\
\multicolumn{2}{c|}{}&\multicolumn{2}{c|}{}&\multicolumn{2}{c|}{}&\multicolumn{2}{c|}{}&\multicolumn{2}{c|}{}\\
\bottomrule[2pt]
\end{tabular}
\vspace{-1.0em}
\end{table*}



\subsection{Answer to RQ 1: Accuracy of Verifying Propagated Vulnerability Code}

In this section, we aim to assess whether TransferFuzz, in comparison with existing methodologies, can accurately validate vulnerabilities in propagated code.

The results presented in Table 2 indicate that TransferFuzz achieves 1.0 of precision and 1.0 of recall. Besides, TransferFuzz successfully validated all 38 propagated vulnerability samples. It can verify vulnerabilities more accurately and comprehensively than existing methods.

\begin{figure}
\centering
    \includegraphics[width=\linewidth]{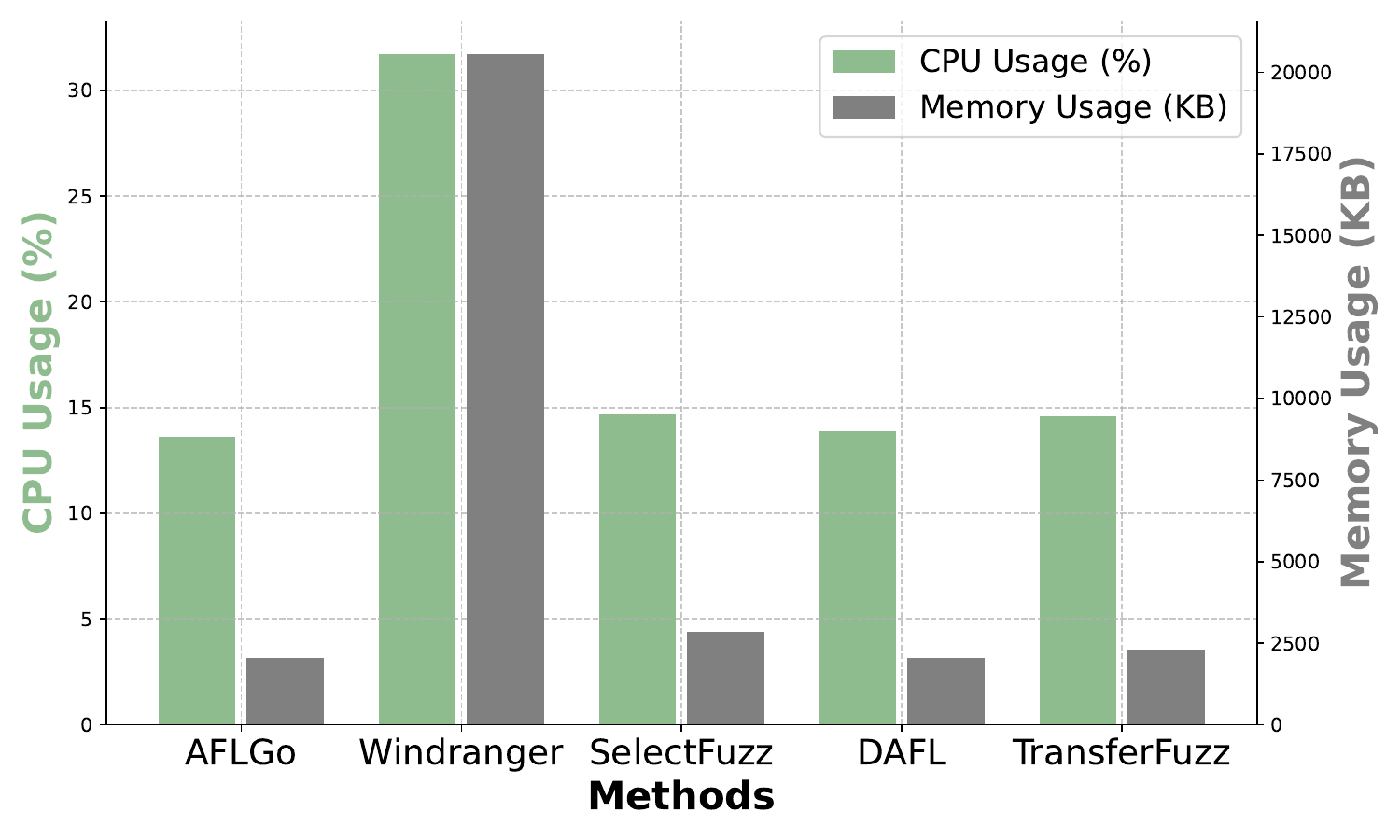}

\vspace{-0.5em}
\caption{CPU and Memory Usage. The green bars represent CPU Usage, while the gray bars indicate Memory Usage (RES).}
\vspace{-0.5em}
\end{figure}

LibAM is another method for achieving a recall of 1.0, as such reuse detection approaches assess code similarity without further validating whether vulnerabilities are triggerable. This results in a significantly low precision of 0.260.

OCTOPOCs is the first method to transfer Proof-of-Concepts (POCs) from basic binaries to target binaries. However, due to its susceptibility to path explosion and the requirement for target binaries to share the same input type as the basic binaries, it effectively only applies to the scenario where the POCs of source binaries and target binaries are consistent, which account for less than half of our datasets.

Other Directed Fuzzing methods demonstrate commendable precision. Yet, their recall rates are significantly lower than that of TransferFuzz. This disparity stems from their limited capacity to validate vulnerabilities with complex logic. Some vulnerabilities are difficult to trigger. TransferFuzz overcomes this challenge by utilizing historical traces derived from POCs in basic binaries.

\vspace{-0.5em}
\begin{tcolorbox}[colback=gray!10,
                  colframe=black,
                  arc=1mm, auto outer arc,
                  boxrule=1.5pt,
                 ]

\textbf{Answering RQ1:} TransferFuzz demonstrates the capability to accurately validate vulnerabilities, achieving the highest precision and recall. TransferFuzz successfully validated 38 propagated vulnerability samples.
\end{tcolorbox}

\subsection{Answer to RQ 2: Efficiency in Triggering Vulnerabilities}

In this section, we evaluate the speed of TransferFuzz compared to existing directed fuzzing methods. Note that we used empty files or simple test files (e.g., basic ELF or PDF files) from the target project as initial seeds for all methods, similar to SelectFuzz \cite{selectfuzz}. Although some POCs of basic binary can directly work on target binaries, they constitute only a small portion of our dataset (Several POCs from LibMing and OCTOPOCs). To ensure a fair evaluation of each fuzzer, we did not take shortcuts by directly using POCs from basic binaries, which is different from the settings in RQ1. We analyze each sample, elucidating the reasons why TransferFuzz outperforms existing methods.

According to Table 3, TransferFuzz ranks as the most rapid method, achieving a speed increase of 2.5 times over the state-of-the-art SelectFuzz. Almost all vulnerabilities are verified within 60 minutes, with over half of them confirmed within 2 minutes. Notably, without using POCs of basic binary, none of the methods, including TransferFuzz, could generate POCs for three vulnerabilities (CVE-2017-18267, CVE-2018-11102, and CVE-2018-20330). This limitation arises because existing methods are prototypes and require additional engineering efforts, such as in-process instrumentation, to adapt to more software. We plan to expand existing fuzzers and undertake these engineering efforts in the future.


Other methods sometimes take several hours to trigger vulnerabilities. For some complex vulnerabilities, such as CVE-2016-4491, CVE-2016-4492, and CVE-2016-6131, they failed to trigger them. Owing to the KBGM algorithm, TransferFuzz can shorten the time cost for triggering from several hours to a  few minutes (e.g., Vul-10). Even for vulnerabilities without complex branches, TransferFuzz significantly speeds up existing methods (e.g., Vul-28 to Vul-35) using the NSA module. Further explanation of the results is provided in Section V-F.

We also recorded the resource consumption of these methods, measuring the maximum CPU usage and memory usage (RES) of TransferFuzz and existing tools during the fuzzing process of each vulnerability. As shown in Figure 5, the resource consumption of TransferFuzz is comparable to AFLGo, SelectFuzz, and DAFL. However, Windranger's memory consumption increased significantly due to the continuous use of taint analysis during the fuzzing process. Despite this, the resource consumption remains acceptable and does not affect the applicability of vulnerability verification.

\begin{tcolorbox}[colback=gray!10,
                  colframe=black,
                  arc=1mm, auto outer arc,
                  boxrule=1.5pt,
                 ]
\textbf{Answering RQ2:} TransferFuzz consistently outperforms existing methods in terms of both speed and the number of vulnerabilities verified.
\end{tcolorbox}

\subsection{Answer to RQ 3: Impact of Different Components}

In this section, we aim to validate the effectiveness of each module within TransferFuzz. We conducted ablation experiments by removing the Key Bytes Guided Mutation strategy, denoted as No KBGM, and the Nested Simulated Annealing algorithm, denoted as No NSA.

In the case of No KBGM, TransferFuzz experienced a notable decline in efficiency due to the lack of key-bytes traces from taint analysis, hindering its ability to bypass essential constraints. This reduction in performance was particularly pronounced for vulnerabilities like CVE-2016-4490, CVE-2016-4491, CVE-2016-4492, and CVE-2016-6131. In contrast, the impact on CVE-2017-11733 was minimal, illustrating that KBGM's effectiveness in extracting key bytes is not universal. Despite these challenges, No KBGM still outperformed the alternative methods in speed.

In the case of No NSA, TransferFuzz experienced a performance decline, although not as pronounced as observed with No KBGM. Despite this, a steady reduction in efficiency was noted since the NSA contributed to enhancements across all examined vulnerabilities.

\begin{tcolorbox}[colback=gray!10,
                  colframe=black,
                  arc=1mm, auto outer arc,
                  boxrule=1.5pt,
                 ]
\textbf{Answering RQ3:} KBGM and NSA both contribute to the performance of TransferFuzz. KBGM can significantly enhance performance under specific circumstances, while NSA consistently improves performance in all cases.
\end{tcolorbox}

\begin{table}[t]
\centering
\caption{Efficiency of different modules of TransferFuzz.}
\label{table2}
\resizebox{\columnwidth}{!}{%
\begin{tabular}{ccccc}
\toprule[2pt]
\multicolumn{1}{c}{\multirow{1}{*}{\textbf{CVE-ID}}} & \multicolumn{1}{c}{\multirow{1}{*}{\textbf{Targets}}} & \multicolumn{1}{c}{\textbf{No KBGM}} & \multicolumn{1}{c}{\textbf{No NSA}} & \multicolumn{1}{c}{\textbf{TransferFuzz}}\\
\hline
\multirow{2}{*}{2016-10095} & thumbnail & 1m59s &  1m2s &  \textbf{0m56s}\\
 & tiffcmp & 3m1s &  2m24s &  \textbf{1m49s}\\
\hline
\multirow{2}{*}{2016-5318} & tiffsplit & 1m54s &  1m31s &  \textbf{1m28s}\\
 & tiffcmp & 2m6s &  1m56s &  \textbf{1m38s}\\
\hline
\multirow{3}{*}{2016-4487} & objdump & 6m32s &  0m9s &  \textbf{0m5s}\\
 & nm-new & 7m45s &  0m51s &  \textbf{0m40s}\\
 & addr2line & 26m51s &  0m56s &  \textbf{0m38s}\\
 \hline
\multirow{3}{*}{2016-4489} & objdump & 14m2s &  0m58s &  \textbf{0m54s}\\
 & nm-new & 11m41s &  0m34s &  \textbf{0m27s}\\
 & addr2line & 47m26s &  1m14s &  \textbf{0m51s}\\
 \hline
\multirow{3}{*}{2016-4490} & objdump & 12m51s &  0m4s &  \textbf{0m4s}\\
 & nm-new & 9m3s &  1m18s &  \textbf{1m3s}\\
 & addr2line & 34m21s &  1m21s &  \textbf{0m53s}\\
 \hline
\multirow{3}{*}{2016-4491} & objdump & N.A. & 6m19s &  \textbf{3m37s}\\
 & nm-new & N.A. & 6m43s &  \textbf{4m15s}\\
 & addr2line & N.A. & 9m12s &  \textbf{7m14s}\\
 \hline
\multirow{3}{*}{2016-4492} & objdump & 19m51s &  \textbf{0m29s} &  0m32s\\
 & nm-new & 25m32s &  0m47s &  \textbf{0m40s}\\
 & addr2line & 1h18m3s(5) & 0m36s &  \textbf{0m17s}\\
 \hline
\multirow{3}{*}{2016-6131} & objdump & N.A. & 0m5s &  \textbf{0m4s}\\
 & nm-new & N.A. & 1m6s &  \textbf{0m56s}\\
 & addr2line & N.A. & 0m47s &  \textbf{0m32s}\\
\hline
\multirow{1}{*}{2017-7303} & objcopy & 1h43m16s(7) & 2h14m(4) & \textbf{1h30m}(6)\\
 \hline
\multirow{4}{*}{2017-11733} & swftophp & 18s &  0m26s &  \textbf{0m6s}\\
 & swftoperl & 0m58s &  0m43s &  \textbf{0m52s}\\
 & swftopython & 0m41s &  0m35s &  \textbf{0m28s}\\
 & swftotcl & 1m12s &  0m37s &  \textbf{0m25s}\\
  \hline
\multirow{4}{*}{2018-8807} & swftocxx & \textbf{18m49s} &  28m13s &  20m2s\\
 & swftoperl & 9m52s &  16m4s &  \textbf{9m21s}\\
 & swftopython & 18m34s &  21m15s &  \textbf{17m2s}\\
 & swftotcl & 22m56s &  23m35s &  \textbf{22m49s}\\
  \hline
\multirow{4}{*}{2018-8962} & swftocxx & 37m25s &  41m38s &  \textbf{33m42s}\\
 & swftoperl & 31m25s &  34m21s &  \textbf{29m15s}\\
 & swftopython & 56m16 & 57m54s &  \textbf{55m20s}\\
 & swftotcl & 51m46s &  \textbf{43m19s} &  46m16s\\
\bottomrule[2pt]
\end{tabular}
}
\vspace{-2.0em}
\end{table}

\subsection{Result Analysis}
In this section, we analyzed the spread of these vulnerabilities and tried to explain the performance of TransferFuzz in these cases.

\textbf{Vul-1 to Vul-4.}
Due to different processing methods for tiff files, the POCs between binaries such as thumbnail and tiffsplit are not adaptable (cann't use basic POC directly in target binaries). However, they all reuse the same vulnerable function and do not perform security checks on the key parameters of the vulnerability, leading to the spread of the vulnerability between these binaries.

These are simple vulnerabilities that all methods can trigger within 10 minutes. Among them, SelectFuzz and TransferFuzz perform the best due to their use of selective instrumentation.

\textbf{Vul-5 to Vul-23.}
These vulnerabilities successfully spread across the \textit{cxxfilt}, \textit{objdump}, \textit{nm-new}, and \textit{addr2line}. However, \textit{cxxfilt} takes a string as input, which is directly used as the parameter of the vulnerable function. In contrast, the other binaries require an ELF file as input. Therefore, it is necessary to construct appropriate input to trigger the vulnerability in target binaries instead of directly using basic POC.

By extracting the key bytes in the POC of \textit{cxxfilt} through KBGM, TransferFuzz only needs to insert these key bytes into specific locations of the initial seed (a simple ELF file). Consequently, whether dealing with complex vulnerabilities (such as CVE-2016-4491 and CVE-2016-4492) or simpler ones (e.g., CVE-2016-4487), TransferFuzz can indiscriminately insert key bytes into random locations in the ELF file to quickly trigger the vulnerabilities. In contrast, other methods need to constantly mutate bytes to meet different conditions and often fail to trigger complex vulnerabilities.

\textbf{Vul-24 to Vul-35.}
For the vulnerabilities in LibMing, although their POCs are universal, we did not use the original POCs during fuzzing to test the capabilities of different fuzzers. Owing to selective instrumentation, both SelectFuzz and TrasnferFuzz triggered these vulnerabilities in a shorter time. Due to the improvement of NSA, TrasnferFuzz is slightly better than SelectFuzz. Unfortunately, DAFL failed to verify CVE-2018-8807 and CVE-2018-8962, which is consistent with the experiments reported in the DAFL paper \cite{dafl}.


\section{Threats to Validity}

The first concern is the effectiveness of TransferFuzz. First, not all vulnerabilities allow for the extraction of key-bytes traces, such as CVE-2017-7303. Despite this, many vulnerabilities do produce such traces, as evidenced by the Binutils dataset where all binaries have key-byte traces, significantly enhancing the performance of our experiments. In addition, some binaries may slow down the fuzzing process due to complex branch constraints or missing key-bytes traces. Nonetheless, TransferFuzz consistently completed verifications within two hours, with over half of the cases resolved in under 1 minutes, which shows the effectiveness of TransferFuzz in verifying vulnerability code propagation.

The second concern is the obstacles to implementation. All processes of TransferFuzz are automated, with only minimal manual work required to write the fuzz driver before fuzzing the target binary, which is a common issue with fuzzing techniques. Although recent works propose automating this task \cite{autofuzz}, these efforts are considered orthogonal to ours. This remaining manual work does not significantly affect the scalability and is much more straightforward for security researchers and developers than manual verification vulnerabilities. Even users familiar with fuzzing technology or the target software can use our tool to verify the target software. 

The third concern is the evaluation dataset. The dataset for evaluation was selected not to intentionally showcase TransferFuzz's strengths but to maintain objectivity. It encompasses all instances from AFLGo and SelectFuzz related to vulnerability code propagation, alongside as many cases as possible from OCTOPOCs (their dataset not being available). Our objective was to employ unbiased datasets for existing methods. However, further empirical studies across a wider range of projects are essential to fully generalize our evaluation results.

\section{Related Works}

\subsection{Code Reuse Detection}

A large number of works have been proposed for code reuse detection. \cite{Code_Similarity_survey1, Code_Similarity_survey2}. First, some work uses constant features to detect code reuse. they identify TPL reuse within the target binary by extracting identical constant features, such as strings \cite{bat, libdx}, function names \cite{osspolice}, and jump tables \cite{b2sfinder}, from both the target binary and the basic binary. In addition, some researchers use binary code similarity detection technology to detect reused code. they involve comparing all the functions of the target binary with those of the basic binary. Subsequently, a predetermined threshold is set to establish whether reuse has occurred. Gitz \cite{gitz} and VIVA \cite{VIVA} use text hashing to represent the function features. BinSim \cite{BinSim} and Bingo \cite{Bingo} use Symbolic execution-based techniques to offer a more robust approach to function similarity detection. Besides, more works \cite{gemini, jtrans} use deep learning technology to learn the semantics of functions. 


All of these techniques can be used as a precursor to TransferFuzz for detecting reused vulnerable code. However they do not further validate these vulnerabilities, and using TransferFuzz can further filter out vulnerability false positives in code reuse detection results.

\subsection{Vulnerability Detection}

Vulnerability detection has long been a popular and significant field in computer science and cybersecurity. Researchers aim to detect vulnerabilities in newly developed code by extracting features from vulnerable functions and determining if these vulnerable functions exist within the target code. 

Early works in this field, such as Bingo \cite{Bingo} and SAFE \cite{safe}, calculated function similarity by directly comparing the vulnerable function with all functions present in the target code. More recent approaches, including MVP \cite{mvp} and VIVA \cite{VIVA}, detect vulnerabilities by employing data stream slicing techniques to identify the presence of vulnerability code and the absence of patch code. Additionally, Fiber \cite{fiber} and PDiff \cite{pdiff} attempt to extract deep patch code semantics by utilizing symbolic execution to ascertain whether the target vulnerable function has been patched or not.

These vulnerability detection efforts likewise do not confirm whether the vulnerability is triggerable. All they do is detect whether a patch exists. Even if the patch doesn't exist, the vulnerability is not necessarily triggerable in a new context.TransferFuzz fills this gap nicely.

\subsection{Directed Grey-box Fuzzing}

Fuzzing, an essential technique in vulnerability discovery for real-world software like parsers \cite{Dfuzzing2019}, network protocols \cite{Pulsar, network_fuzzing}, web browsers \cite{CodeAlchemist, Superion}, mobile apps \cite{Westworld}, and OS kernels \cite{NTFuzz, Razzer, kafl}, has seen significant advancements, particularly in Grey-box fuzzing \cite{bohme2020boosting, WEIZZ, fuzz_survey}.

Directed Grey-box Fuzzing (DGF) addresses the persistent challenge of generating test cases to expose specific program bugs. AFLGo \cite{aflgo}, one of the earliest directed fuzzers, prioritizes test case generation by favoring nodes closer to the target node within the Control-Flow Graph (CFG). Hawkeye \cite{hawkeye} and WindRanger \cite{WindRanger} build upon the AFLGo approach, enhancing it to provide more informative feedback for precise distance computation. 


TransferFuzz is the first to use historical fuzzing data across software to guide new fuzz, which greatly improves the capabilities of the Directed Fuzzer.

\section{Conclusion}

Code reuse in software development commonly leads to the propagation of vulnerability code. Existing methods primarily focus on detecting the presence of vulnerable code in target software, without automatically verifying whether the vulnerability poses a real threat in the new context. This deficiency results in many false positives. In this paper, we introduce TransferFuzz, a novel vulnerability verification framework, to verify whether vulnerabilities propagated through code reuse are actually vulnerable in new software.
Our experiments show that TransferFuzz rapidly detects the exact software scope affected by the vulnerability and generates Proofs-of-Concept to provide strong evidence for the propagation of the vulnerability. Furthermore, TransferFuzz expands the impacted software scope of ten vulnerabilities in the CVE report from 15 binaries to 53, underscoring its effectiveness in verifying propagated vulnerability code.

\section{Acknowledge}
We appreciate all the anonymous reviewers for their invaluable comments and suggestions. This work is partly supported by National Key Research and Development Program of China (No.2022YFB3103904). Any opinions, findings and conclusions in this paper are those of the authors and do not necessarily reflect the views of the funding agencies.




\bibliographystyle{IEEEtran}
\bibliography{software}

\vspace{12pt}

\end{document}